\documentclass[10pt,prc,aps]{revtex4}                                   
\usepackage{graphicx}
\tighten
\begin{document}
\draft
\title{Probing the sensitivity of the total nucleus-nucleus                           
reaction cross section at intermediate energies                                                
to medium effects and isospin asymmetries.                          
}              
\author{            
 Francesca Sammarruca and Larz White}      
\affiliation{                 
 Physics Department, University of Idaho, Moscow, ID 83844-0903, U.S.A  } 
\date{\today} 
\email{fsammarr@uidaho.edu}
\begin{abstract}
This paper presents reaction cross section predictions. These predictions 
are the result of a continuous pipeline which originates from a microscopic nuclear interaction. Density parameters and 
effective nucleon-nucleon cross sections (both involved in the reaction calculations) 
are by-products of the same equation of state. 
First, we perform tests of sensitivity to medium effects using reactions involving $^{208}$Pb,
a stable but weakly isospin-asymmetric nucleus. 
We also  show predictions for collisions of some neutron-rich isotopes of Calcium and Argon. 
We observe significant sensitivity of the reaction cross section to medium effects 
but very weak sensitivity to inclusion of isospin asymmetry in the effective nucleon-nucleon 
cross sections. 
\end{abstract}
\maketitle

\section{Introduction}

Collisions of neutron-rich ions can be exploited to investigate isospin asymmetries of nuclear matter. 
This is a topic of great current interest which impacts both terrestrial nuclear physics and the 
physics of neutron stars. 
The on-going and up-coming experimental programs using radioactive nuclear beams are expected to improve our
knowledge of nuclear properties away from the stability line, making predictions of neutron-rich systems 
especially interesting and timely. 

Interaction cross section  measurements can provide valuable information on the structure of exotic nuclei. 
In this paper, we will be concerned specifically with the nucleus-nucleus total reaction cross section and its 
sensitivity to medium effects and 
to the inclusion of isospin asymmetries.                                                                    
By isospin asymmetries we mean: 
1) Separating neutron and proton densities in the  target and projectile nuclei,                           
as opposed to just considering total {\it nucleon} densites (something which is often done for mildly isospin-asymmetric nuclei); and 2)  
{\it Medium-induced} differences between collisions of different types of nucleons in 
the presence of unequal concentrations of protons and neutrons. (Notice that we are not referring to 
the well-known differences between $pp/nn$ and $np$ cross sections.) 
Intermediate energies are best for our purpose, as both         
the model approximations we apply and the expectation of sensitivity to medium effects are reasonably justified. 

Isospin-dependent dynamics can be studied through a variety of 
reactions/phenomena related to the transport of protons and neutrons.
A review of isospin effects in heavy-ion reactions which may be particularly
suitable to probe the nuclear symmetry energy can be found in Ref.~\cite{Li08}. 
Furthermore, fusion dynamics in heavy-ion collisions close to the 
Coulomb barrier is sensitive to the incident energy and the 
neutron-to-proton ratio in the neck region of the colliding systems \cite{Feng}. 
More recently, isospin and symmetry energy effects on the 
competition between fusion and break-up reactions in low-energy 
(about 10A MeV) heavy-ion collisions have been studied within a stochastic 
mean-field approach \cite{Rizzo}. 
When exotic 
beams are employed, these mechanisms are expected to be sensitive to
the isovector part of the nuclear potential, thus yielding information
on the symmetry energy. In passing, we note that the mean field 
used in calculations such as those from Ref.~\cite{Rizzo} is often
derived from phenomenological forces. We hope that these very 
comprehensive reaction studies will consider including predictions
from microscopic many-body calculations as part of their input. 

Although a rather global observable as compared to the potentially 
more ``selective" mechanisms mentioned above, 
the reaction cross section has the potential to reveal sensitivities to effects
induced by isospin asymmetry in the medium, particularly                  
when written in such a way as to discriminate
between protons and neutrons (see Section {\bf IIA}, Eq.~(8)).              
It is then interesting to explore
to which extent this is true.                     

This paper is                                           
organized as follows. In Section {\bf II}, we review some formal aspects of reaction cross
section calculations (within the optical limit of the Glauber model (GM) \cite{Glauber}) and describe the main input, namely the nucleon-nucleon (NN) collisions and the nuclear densities. 
In our case, 
both of these ingredients are obtained from, or are closely related to, our microscopic equation of state (EoS), making
all aspects of our calculation internally consistent with one underlying nuclear interaction. 
We will discuss how medium effects and isospin asymmetries are introduced in the various elements of our 
calculation.  Within Section {\bf II},            
we will look at the reaction cross section for collisions of stable nuclei and perform some 
sensitivity studies. We have chosen collisions of 
$^{40}$Ca on      
$^{208}$Pb as our initial ``laboratory". Although not an exotic one, $^{208}$Pb is an interesting nucleus to consider, since very accurate
measurements of its neutron skin are expected to come from the electroweak physics program at the Jefferson Laboratory. 
(To the best of our knowledge, such value is not yet available, but we understand that the JLAB target 
uncertainty should be very small.) 

Several authors have confronted reaction cross sections with neutron-rich isotopes (see, for instance,
Ref.~\cite{Abu08} for Carbon isotopes incident on a proton). 
In Section {\bf III} we will look at some reaction cross section data for 
 collisions involving neutron-rich isotopes of Calcium and Argon 
 at energy/nucleon between 50 and 70 MeV \cite{Licot}. 
 For that purpose, we need neutron and proton densities for the 
nuclides under consideration. Again, we obtain density distributions from our equation of state                        
and then move to predictions of reaction cross sections. 

Our conclusions and future outlook are summarized in Section IV.

\section{The reaction cross section} 
\subsection{Review of some formal aspects} 

We operate within the optical limit of the Glauber model \cite{Glauber}. 
The reaction cross section is written, in terms 
of the impact parameter $b$, as
\begin{equation}
\sigma_R = 2\pi \int_0^{\infty} (1-T(b))bdb  \; ,  
\end{equation}
with $T(b)$ the {\it nuclear transparency}, defined as 
\begin{equation}
T(b) = e^{-\int P (b,z) dz }                 \; , 
\end{equation}
where the {\it thickness function} is given by \cite{Karol} 
\begin{equation}
 P (b,z) dz  = \bar {\sigma}(E) \rho_T(b,z)\rho_P(b,z) dz \; .
\end{equation}
Namely, the thickness function 
is the product of the averaged NN cross section and the overlap
integral of the target and projectile densities, $\rho_T$ and $\rho_P$.                 
The averaged NN cross section is defined as 
\begin{equation}
\bar {\sigma}(E) = \Big ( \frac{Z_T}{A_T}\frac{Z_P}{A_P} \Big ) \sigma_{pp}(E) + 
 \Big (                                     
 \frac{N_T}{A_T} \frac{N_P}{A_P} \Big ) \sigma_{nn}(E) + 
\Big ( \frac{Z_T}{A_T} \frac{N_P}{A_P} + 
 \frac{N_T}{A_T} \frac{Z_P}{A_P} \Big ) \sigma_{np}(E)  \; , 
\end{equation}
where $Z_{T(P)}$, 
 $N_{T(P)}$, and 
 $A_{T(P)}$ are the proton, neutron, and mass numbers of the target(projectile) nucleus, and 
$\sigma_{ii}$($\sigma_{ij}$) are the elementary cross sections for scattering of identical(non-identical)
nucleons. 

More generally, and accounting for finite range nuclear interactions \cite{BBS}, one writes
the transparency as 
\begin{equation}
T(b)  =exp \large (- \bar {\sigma}(E) \int d^2r_T \int d^2 r_P f(|{\vec r}_P-{\vec r}_T|)
\rho_z^{(T)}(|{\vec r_T}|) \rho_z^{(P)}(|{\vec r_P} - {\vec b}|) \large ) \; , 
\end{equation}
where the $z$-integrated densities of the target and projectile nuclei are defined as
\begin{equation}
\rho_z^{(T)}(|{\vec r_T}|) = \int dz \rho^{(T)} ((r_T^2 + z^2)^{1/2})             
\end{equation}
and 
\begin{equation}
\rho_z^{(P)}(|{\vec r_P}-{\vec b}|) = \int dz \rho^{(P)} ((|{\vec r_P}-{\vec b}|^2 + z^2)^{1/2}) \; .        
\end{equation}
In these expressions       
${\vec r}_P$ and ${\vec r}_T$ are {\it transverse} coordinates, defined in the plane perpendicular to the 
beam axis. 

Typically, the range function,                                  
$f(|{\vec r}_T-{\vec r}_P|)$, which is normalized to unity, is taken to be of Gaussian shape with an interaction range of about 0.6-1.0 fm 
 \cite{BBS}. Taking the range function to be a delta function, 
$\delta({\vec r_T} - {\vec r_P})$, amounts to adopting the zero-range approximation~\cite{Karol}.                      

The formulas above can be understood in terms of interaction probability between density elements
in the target and projectile nuclei. Those 
density elements consist of ``tubes" parallel to the beam axis (the $z$-axis) and whose position is specified by 
the vectors ${\vec r_T}$ and 
 ${\vec r_P}$. The use of z-integrated 
densities reduces an otherwise six-dimensional integral (similar to the double-folding of an effective nuclear interation) to a four-dimensional one, which we calculate numerically. Alternatively, the integral can be rendered separable by 
Fourier transformation into momentum space \cite{Satch}.

Notice that, when properly discriminating between proton and neutron densities (beyond the $Z/A$ and $N/A$ factors
which appear in Eq.~(4)), the transparency function is written as 
\begin{equation}
T(b)  =exp \large (- \int d^2r_T \int d^2 r_P f(|{\vec r}_T-{\vec r}_P|)    
 \sum_{i=n,p}\sum_{j=n,p}\sigma_{ij}
\rho_{z,i}^{(T)}({\vec r_T}) \rho_{z,j}^{(P)}(|{\vec r_P} - {\vec b}|) \large ) \; , 
\end{equation}
where the summations extend over neutron and proton densities of target and projectile. 
In the next section, we will discuss the inclusion of isospin asymmetry through the densities as well as
the elementary cross sections. 

A first-order Coulomb correction can be included by 
replacing the impact parameter with the (classical) distance of closest approach in the Coulomb field 
\cite{Vries,Shukla}. That is, in Eq.~(5) one replaces the impact parameter with 
\begin{equation}
 d = \frac{a}{2} + \sqrt{\Big (\frac{a}{2} \Big )^2 + b^2} \; , 
\end{equation}
 where $ a$ is the distance of closest approach when $b=0$. 
We apply such trajectory modification here. 

Some comments are in place concerning the validity of the 
Glauber model in the energy range to be considered here, 
particularly the lower end of that range, which is 50 MeV/nucleon. 
The basic assumption of the Glauber model is the description of the
relative motion of target and projectile in terms of straight 
trajectories. For low energy, the eikonal trajectories should be 
modified to account for both the Coulomb field and the 
nuclear field, with                                         
the nuclear field effect
tending to move the reaction cross section closer to the 
Glauber model results.                                                         
In Ref.~\cite{Shukla}, it is shown that 
the most significant effect at low and intermediate energies           
(from the Coulomb barrier to 50 times the Coulomb barrier) 
is the one due to the Coulomb field. In fact, close to the Coulomb 
barrier, the Coulomb-modified GM and the Coulomb plus nuclear modified GM
appear indistinguishable. 
For a system such as $^{40}$Ca + $^{208}$Pb, which we will confront next,
50 MeV/nucleon is about 10 times the Coulomb barrier (in terms of 
center-of-mass energy), and the results of Ref.~\cite{Shukla} gives
us some confidence that a (Coulomb-modified) GM is reasonably valid for our 
purposes. 
Kox {\it et al.} \cite{Kox} included effects        
from the nuclear potential to compare modified GM predictions with their measurements 
(in the range of 10-300 MeV/nucleon). The potential was taken to be 
the real part of an optical potential extracted from analysis of 
elastic scattering data for similar reaction systems.                    
They note that nuclear field effects 
can counterbalance
the Coulomb repulsion in a significant way mostly 
for the lightest projectiles \cite{BS80} (one reason why we prefer to work      
with medium-heavy nuclei).

Another mechanism that goes beyond the Glauber model is Pauli blocking,             
which is included in 
our in-medium NN cross sections (see below). Note, further, that our in-medium cross sections
also contain nuclear mean field effects, 
being modified by the average nuclear matter
potential. 

Weighing all the considerations above, we remain well aware that
several modifications of the Glauber model may be important at the lower
energies, especially when a detailed comparison with the data is the 
main objective. Here, on the other hand, we focus mostly on exploring sensitivities. 


\begin{figure}[!t] 
\centering 
\vspace*{-1.0cm}
\hspace*{-1.0cm}
\scalebox{0.30}{\includegraphics{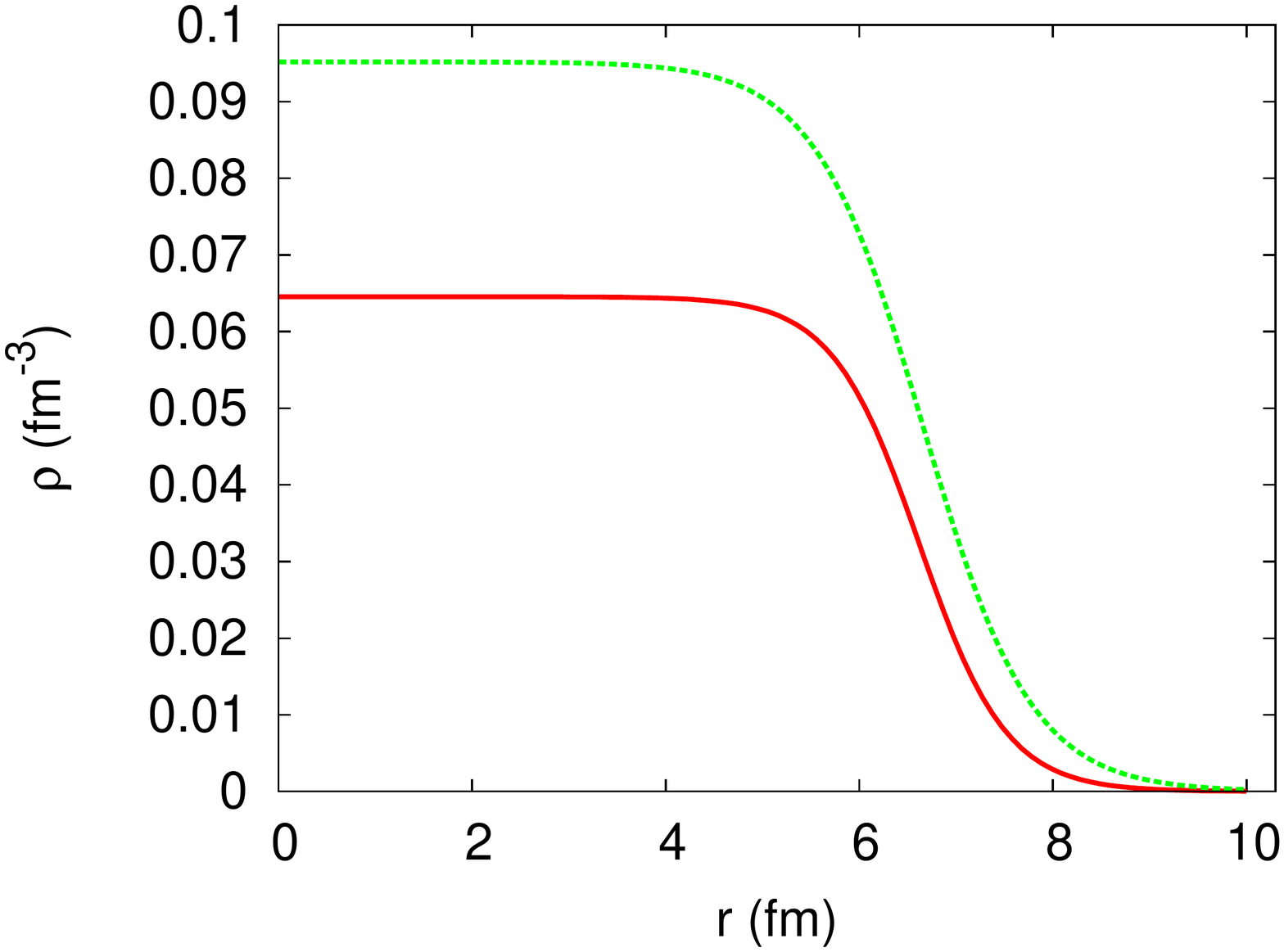}} 
\scalebox{0.30}{\includegraphics{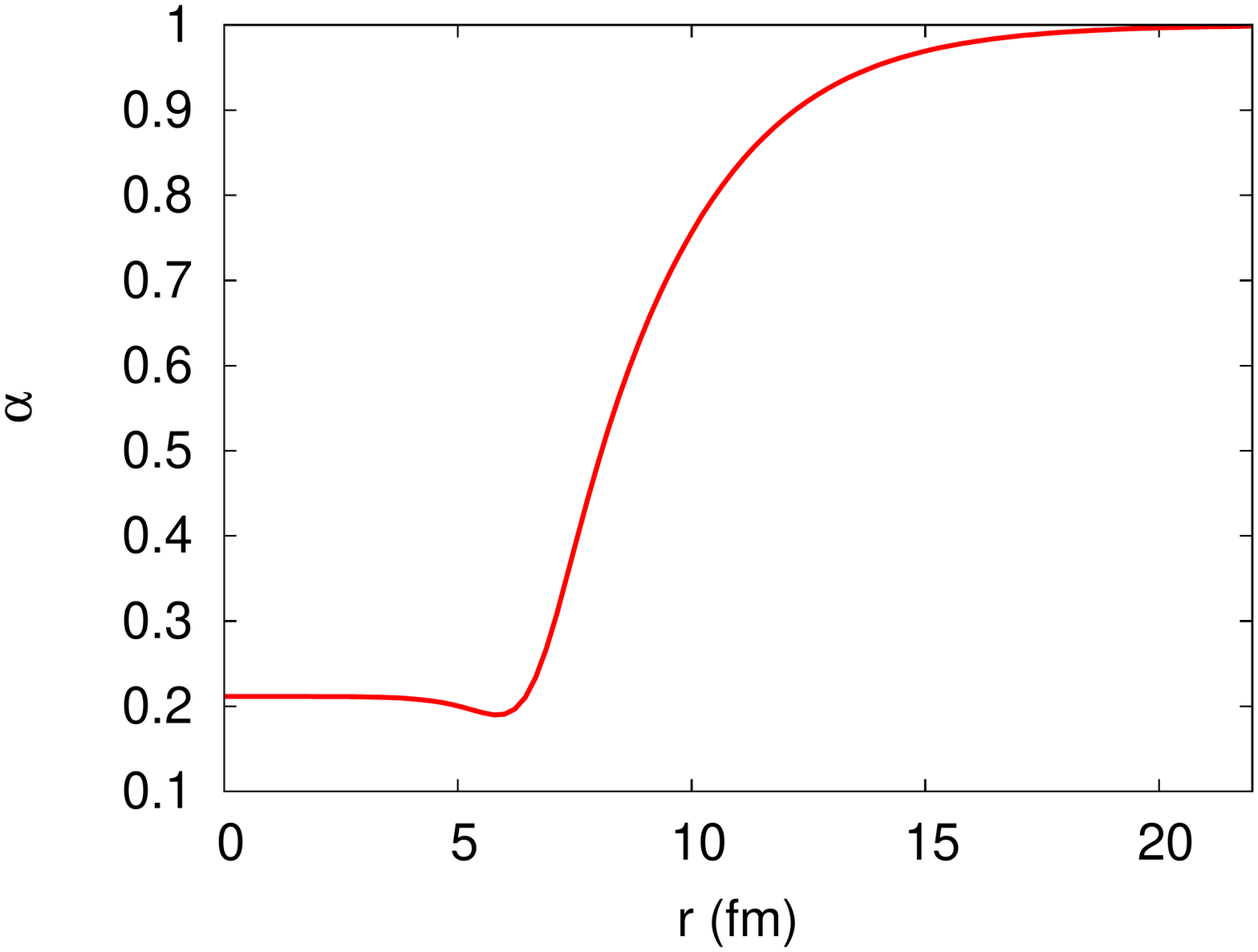}} 
\vspace*{-0.5cm}
\caption{(color online)                                        
Left frame: 
Two-parameter Fermi functions used to describe the neutron (dotted green curve) and the proton (solid red curve) densities in $^{208}$Pb.  
Right frame: the neutron excess parameter, $\alpha$, 
corresponding to the densities given on the left.            
} 
\label{one}
\end{figure}

\begin{table}                
\centering \caption                                                    
{ Point and charge r.m.s. radii (in units of fermi) predicted through our equation of state for $^{40}$Ca and              
$^{208}$Pb. The second and third column display the proton and neutron point radii. The fourth and fifth columns show the charge radius and its empirical value. The values in the second half of the Table are obtained ignoring differencs between proton and neutron distributions. See 
text for details. 
} 
\vspace{5mm}
\begin{tabular}{|c|c|c|c|c|}
\hline
Nucleus  &$ <r_{r.m.s}>_p $ &$ <r_{r.m.s.}>_n$ &  $<r_{r.m.s.}>_{ch}$ & $<r_{r.m.s.}>_{ch}^{exp}$ \\            
\hline
       $^{208}$Pb & 5.39 & 5.56   &5.45   & 5.52   \\ 
       $^{40}$Ca & 3.43 & 3.39   &3.52   &3.48   \\ 
\hline
       $^{208}$Pb &5.45  & 5.45   &5.51   & 5.52   \\ 
       $^{40}$Ca & 3.41 & 3.41   &3.50   & 3.48  \\ 
\hline
\end{tabular}
\end{table}
\subsection{The proton and neutron densities} 

Our density functions are theoretical predictions obtained directly as point densities  
with the method described in Refs.~\cite{FS10,SL09}. Our EoS is                        
derived within the Dirac-Brueckner-Hartree-Fock (DBHF) approach to nuclear matter together with 
the Bonn B NN potential \cite{bonn}. 
We write an energy functional based on the mass formula, where the ``volume term" contains the 
energy per particle as a function of density (namely, the EoS). The density functions that 
appear in the various integral terms of the mass formula are parametrized as two-parameter
Fermi functions (best suited for medium to heavy nuclei),                           
\begin{equation}
\rho(r) = \frac{\rho_0}{1 + e^{(r-r_0)/c}} \; , 
\end{equation}
and the parameters $r_0$ and $c$ are then determined from energy minimization. In this way, when the asymmetric matter
EoS is used in the mass formula, we can extract                        
parameters of both neutron and proton densities, as we have done previously for the purpose of predicting the neutron skin  of $^{208}$Pb \cite{FS10,SL09}. The proton and neutron densities in $^{208}$Pb thus calculated are shown in the left frame of Fig.~1. The
corresponding neutron skin is equal to 0.169 fm.                                        
On the right side of the same figure we show the predicted isospin asymmetry parameter in $^{208}$Pb, $\alpha=(\rho_n-\rho_p)/(\rho_n+\rho_p)$, as 
a function of the radial coordinate from the center of the nucleus.                                          
 The asymmetry parameter is close to 0.2 up to
a radial distance of about 5 fm, but it can reach values up to 0.7 in the skin. 

In Table I, we show our predicted r.m.s. radii for 
       $^{208}$Pb and                                    
       $^{40}$Ca. The values displayed in the upper half of the Table are obtained with our asymmetric matter
EoS. For
       $^{40}$Ca our calculation predicts  a small but negative skin, an indication that the neutrons
form a slightly more compact system than the protons. 
The values in the lower half of the Table are obtained imposing that the density profile is the same for neutrons 
and protons. Consistent with that assumption (which clearly results into a neutron skin equal to zero), we use the symmetric matter equation of state in the energy
functional when searching for those parameters. 

We now move to a discussion of Fig.~2. 
The solid line shows the predictions we obtain ignoring differences between proton and neutron 
density distributions,                             
which are then scaled with 
 $Z/A$ and $N/A$ factors as in Eq.~(4),                                                         
 whereas the dotted line is obtained with 
the neutron and proton densities shown in Fig.~1.                                   
Typical differences between the solid and the dotted lines amount to about 2-3\%. 
The dash-dotted line in Fig.~2               
is obtained using neutron and proton density functions derived from a different EoS, as we explain next.
In Ref.~\cite{FS10} we compared a family of EoS based on the Brueckner-Hartree-Fock approach implemented
with three-body forces (TBF), on the one hand, and our DBHF-based EoS, on the other. 
In particular, we compared predictions of the 
symmetry pressure, proportional to the derivative of the symmetry energy, and the (strongly correlated) neutron skin
in $^{208}$Pb.             
The EoS employed for the dash-dotted line in Fig.~2 is the one referred to as UIX in Ref.~\cite{FS10}, 
which utilizes the Argonne V18 potential \cite{V18} and the phenomenological Urbana TBF \cite{UIX}. 
It is considerably more repulsive than our EoS and predicts a larger neutron skin for $^{208}$Pb. According to our calculation,
that value is 
0.226 fm {\it vs} 0.169 fm (our prediction), a 33\% increase. Nevertheless, the (EoS-induced) differences between the dotted and the dash-dotted 
lines in Fig.~2 are of the order of 4-5\%, 
implying that data precision at this level would  
be necessary to distinguish between the two EoS applied for this test.                    
On the other hand, we must keep in mind that                                             
in a different system, such as a                                                    
nucleus with a very large skin or halo, the model dependence displayed in Fig.~2 may be more significant 
and may have the potential to help discriminating among different EoS.

\begin{figure}[!t] 
\centering 
\vspace*{-1.0cm}
\hspace*{-1.0cm}
\scalebox{0.35}{\includegraphics{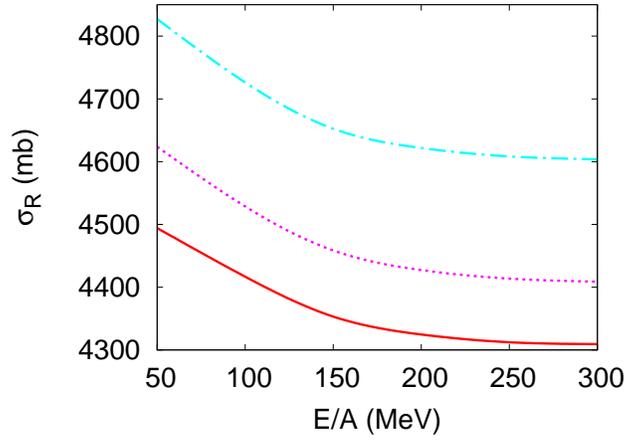}} 
\vspace*{-0.5cm}
\caption{(color online)                                        
Total reaction cross section for $^{40}$Ca on $^{208}$Pb as a function of the energy/nucleon.   
The dotted (pink) line is obtained with proton and neutron densities leading to the predictions in the 
upper half of Table I, whereas the solid (red) curve corresponds to the predictions in the lower half
of Table I. The dash-dotted (blue) line is obtained with density functions derived from a different EoS, as  
explained in the text. 
Free-space NN cross sections are employed. 
} 
\label{two}
\end{figure}

\begin{figure}[!t] 
\centering 
\vspace*{-1.0cm}
\hspace*{-1.0cm}
\scalebox{0.30}{\includegraphics{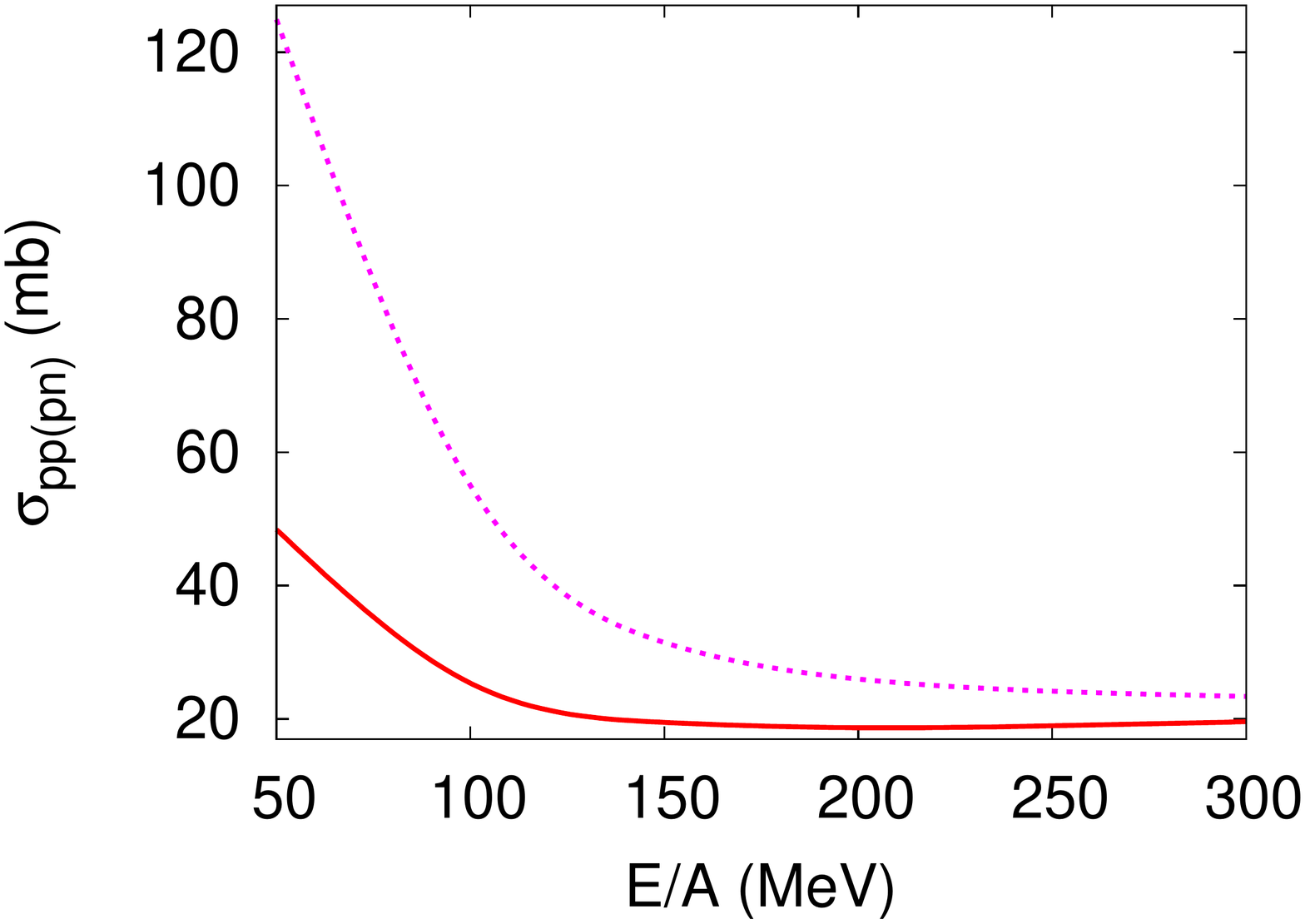}} 
\scalebox{0.30}{\includegraphics{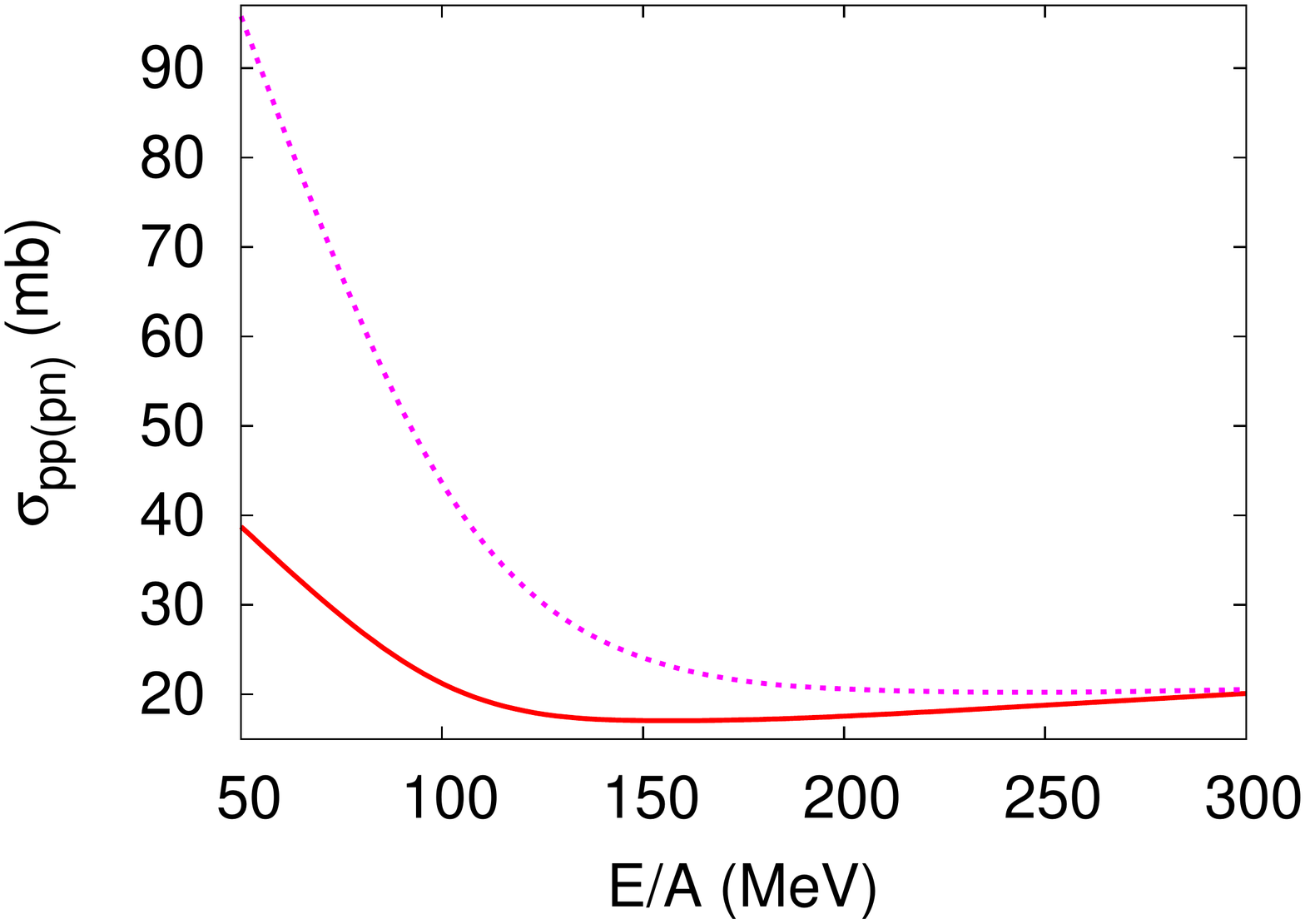}} 
\vspace*{-0.5cm}
\caption{(color online)                                        
Our in-medium NN cross sections                                                                   
in symmetric matter with a Fermi momentum equal to 1.1 fm$^{-1}$ (left side)  
and 1.3 fm$^{-1}$ (right side). Dotted (pink) line: $np$; solid (red) line: identical nucleons. 
} 
\label{three}
\end{figure}

\begin{figure}[!t] 
\centering 
\vspace*{-1.0cm}
\hspace*{-1.0cm}
\scalebox{0.30}{\includegraphics{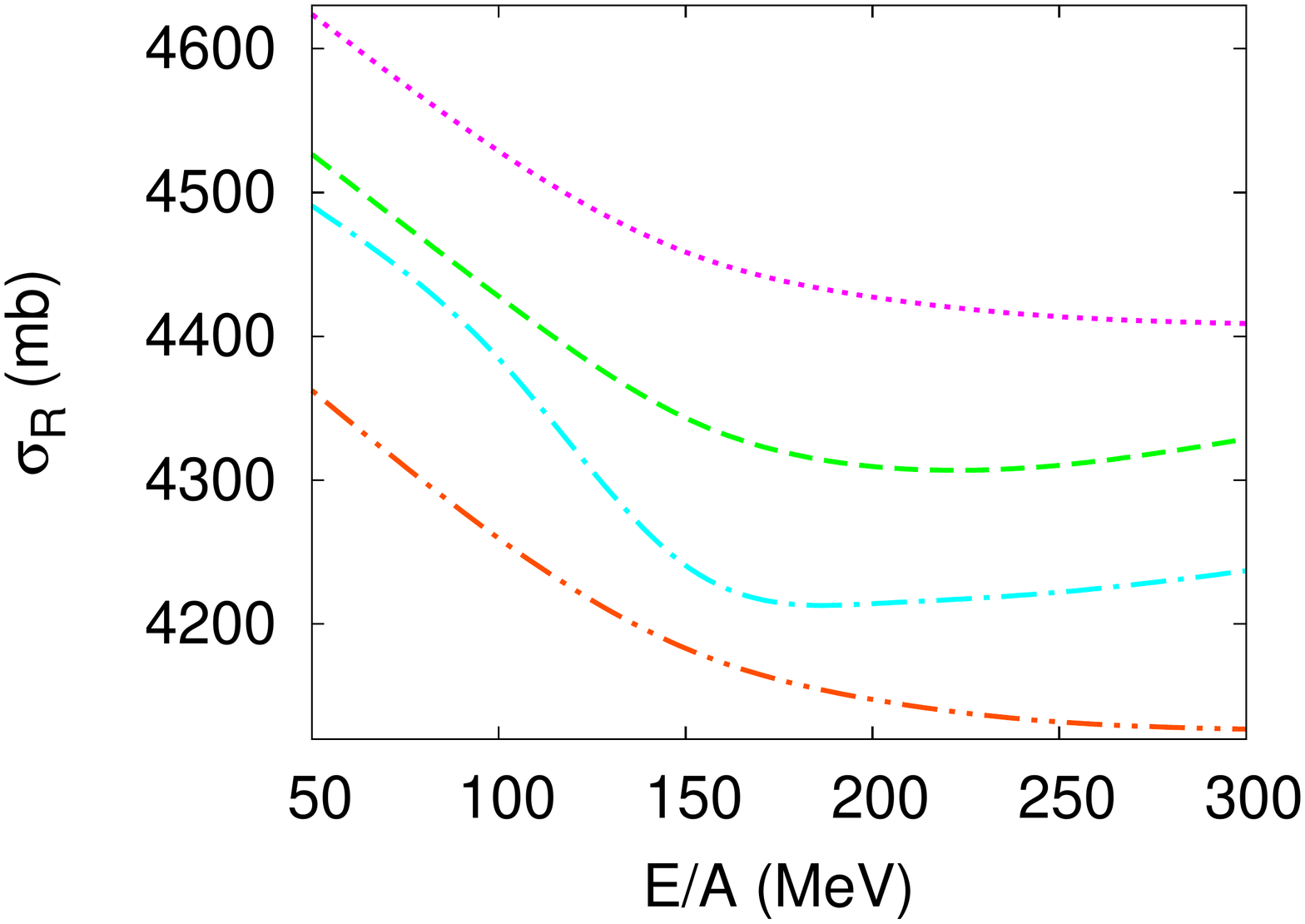}} 
\scalebox{0.30}{\includegraphics{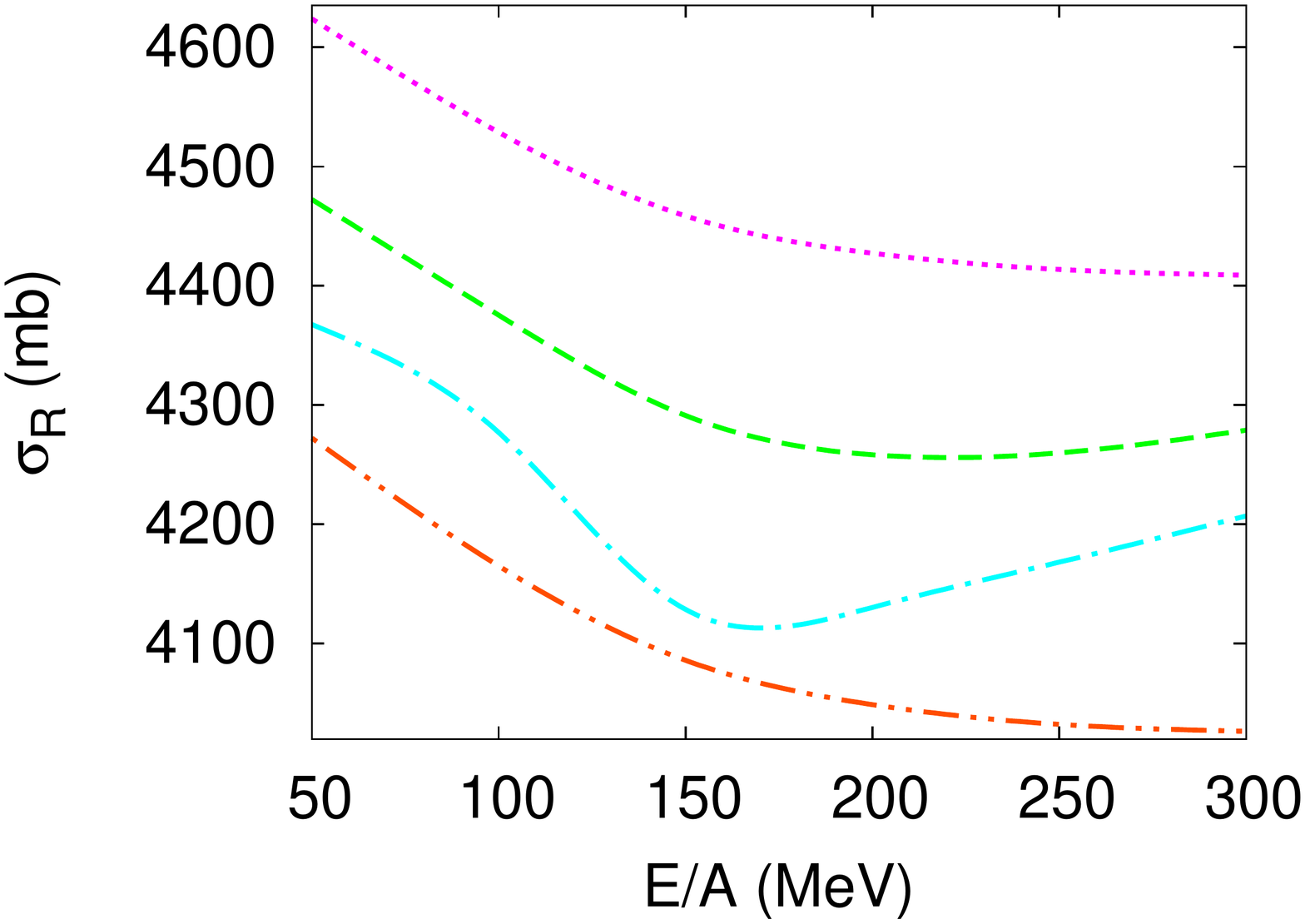}} 
\vspace*{-0.5cm}
\caption{(color online)                                        
Total reaction cross section for $^{40}$Ca on $^{208}$Pb as a function of the energy/nucleon.   
The dotted (pink) line employs free-space NN cross sections. The other three predictions make use of
different models for the calculation of medium effects, as detailed in the text. 
An average density equal to 0.089 fm$^{-3}$ (0.148 $fm^{-3}$) is assumed in the left (right) panel. 
} 
\label{four}
\end{figure}

\subsection{The nucleon-nucleon cross sections: medium and isospin asymmetry effects} 

In previous papers \cite{SK06,Catania} we have confronted the issue of microscopic in-medium NN cross sections. 
We discussed how the very definition of in-medium cross sections 
can be ambiguous, because the scenario of two particles scattering in the medium requires,                           
 at the very least, kinematical choices that are not unique.                                                        
Furthermore, how to properly include
Pauli blocking                                                                                       
has to be considered carefully \cite{SK06,Bert10}. 

The effective NN cross sections we apply in this work are shown in Fig.~3 at two different densities of 
symmetric matter. 
They include modifications of the microscopic G-matrix through Pauli blocking 
of the intermediate states and modifications of the single-particle energy by the nuclear matter mean field, 
parametrized through the nucleon effective mass. The latter is a direct by-product of the EoS        
calculation. 

In a much simpler approach, one makes the assumption that the transition matrix in the medium is approximately 
the same as the one in vacuum and that medium effects come in only through the use of effective masses
in the phase space factor \cite{Pand,Pers,Li05}. In that case, in-medium cross sections are scaled (relative to their 
value in vacuum) as the square of the ratio of the (reduced) masses. Namely,
\begin{equation}
\sigma^*_{NN}/\sigma_{NN} = (\mu^*_{NN}/\mu_{NN})^2 \; , 
\end{equation} 
where $\sigma^*$ stands for the cross section at non-zero density. 
In summary, the issue of how to define and calculate microscopic in-medium NN cross sections remains an 
interesting and unsettled one. 

We will now proceed to explore sensitivity of the reaction cross section with respect to the model 
adopted to describe medium effects on the NN cross sections.

In Fig.~4, we show again the reaction cross section for 
$^{40}$Ca on $^{208}$Pb as a function of the energy/nucleon.                                           
The dotted (pink) line is the same as the dotted line in Fig.~2 and employs free-space NN
cross sections. 
All other predictions include medium effects calculated in {\it symmetric} nuclear matter at 
some average density of 0.089 fm$^{-3}$ (left frame) and                                                      
0.148 fm$^{-3}$ (right frame).                                                         
 The dash-dotted (blue) line employs our microscopic in-medium NN cross sections as shown in Fig.~3 
(method I).        
As mentioned above, 
they are the result of DBHF G-matrix calculations which include Pauli blocking of the intermediate states
and (self-consistent) modification of the nucleon mass through dispersive effects on the single-particle
energy. 
The predictions shown by the dashed (green) line make use of 
in-medium NN cross sections derived from the phenomenological formula of Ref.~\cite{sigmed} (method II).               
The latter combines the energy dependence
of free space cross sections with the density dependence from the microscopic model 
of Refs.~\cite{LM,LM2}, although we found that the agreement with the latter is only approximate. 
At this point we note that the calculations of          
Refs.~\cite{LM,LM2} are also based on the                                                         
DBHF approach to nuclear matter, but medium effects are applied to the (real) K-matrix while we
work with the (complex) G-matrix. We found that it is important to do so (even below the inelastic threshold, 
where of course the two methods would yield identical predictions in free space), 
because application of Pauli blocking, by removing energetically open channels, amounts to violation 
of free-space unitarity. On the other hand, calculation of NN phase shifts from the 
real K-matrix imposes free-space unitarity. We believe this is the reason why there is 
a non-negligeable difference between our predictions and those of Refs.\cite{LM,LM2}. 
Finally, 
the dash-double-dotted (orange) line is obtained using NN cross sections from the mass scaling formula, Eq.~(11), 
where we have used                                                 
 our dynamically generated effective masses (method III).

The strongest medium effects are seen with the simplest approach, namely method III.                 
Also, medium effects are non-linear with density.
It is interesting to observe that the predictions obtained with the microscopic NN cross sections show
more structure in their energy dependence. The slow rise at the higher energies (see left panel of Fig.~4)
is very similar to the one displayed by the dashed line (we recall that the latter uses NN in-medium cross sections
which, to a certain extent, 
simulate a microscopic DBHF model). Such rise is more pronounced at the higher density 
(see right side of the figure) and is mostly due to enhancement of the in-medium $pp$ cross sections, 
a phenomenon that has been observed by other authors as well \cite{Catania,LM,LM2}. 

The largest differences seen in Fig.~4 (relative to the free-space results) amount to about 6\% at the lower
density (left frame) 
and approximately 8-9\% at the higher density (right frame).                                                  
We conclude that medium effects can be significant. 
(Of course 
the densities involved in a particular collision are functions of space, and so should be the NN in-medium 
cross sections. With 
our sensitivity tests we have certainly 
covered a typical density range probed by these reactions.)

We now move on to effects from neutron/proton asymmetry in the NN collisions. 
Our           
DBHF calculation of the asymmetric matter EoS and related quantities (such as the symmetry potential
and the effective masses for neutrons and protons) is described in details in Ref.~\cite{FS10}. 
For a given total density, we allow the population of one nucleon species to increase. The presence of
two different Fermi levels 
results into self-consistent potentials, and thus effective masses, which are different for 
neutrons and protons. 
As a consequence, 
in-medium cross sections for identical nucleons ($pp$ or $nn$) are different from 
each other. (It should be clear that we are referring to strong interaction cross sections and to differences
that are unrelated to charge.) 
We also explored isospin effects on the in-medium $pn$ cross section and found them entirely negligeable, due to 
the fact that the two different nucleon species are oppositely impacted by those effects as the 
population of one species increases. 

To include isospin asymmetry within 
 method III, we simply apply the appropriate effective masses (this time, different for protons and neutrons) in 
Eq.~(11). 
(The formula previously referred to as ``method II" does not incorporate this kind of isospin          
degree of freedom and thus is not included in the discussion of Fig.~5.) 

These NN cross sections, now energy, density, and $\alpha$ dependent, 
are then inserted in Eq.~(8). 
Figure 5 shows the total reaction cross sections at 100 MeV (in terms of E/A) and a fixed moderate density
($\rho$=0.068 fm$^{-3}$)
as a function of the neutron excess parameter as defined in Section {\bf IIB}. More precisely, each value of $\alpha$ represents some 
average neutron/proton asymmetry that is being attributed to the colliding system for the purpose of 
sensitivity testing. Clearly, the slope of the two lines shown in the figure is insignificant, especially
compared to the differences between the two sets of predictions. 
Thus, sensitivity to isospin dependence of the NN cross sections can be disregarded in this observable, 
an observation that may be of practical relevance. 

\begin{figure}[!t] 
\centering 
\vspace*{-1.0cm}
\hspace*{-1.0cm}
\scalebox{0.30}{\includegraphics{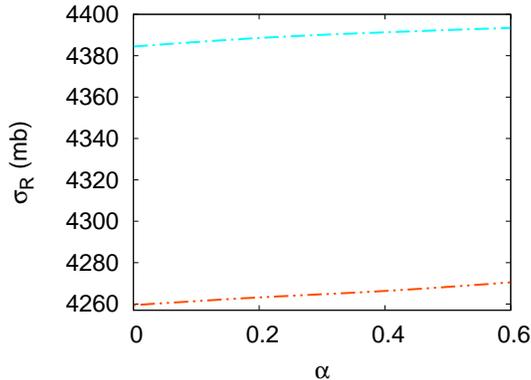}} 
\vspace*{-0.5cm}
\caption{(color online)                                        
Total reaction cross section for $^{40}$Ca on $^{208}$Pb at E/A=100 MeV. The effective NN cross sections
are calculated with method I (dash-dotted, blue) and method III (dash-double-dotted, orange), assuming 
an average density of 0.068 fm$^{-3}$ and including isospin asymmetry as explained in the text. 
} 
\label{five}
\end{figure}

\section{Reaction cross section with neutron-rich isotopes} 
In this section, we will consider reactions with some neutron-rich isotopes of Calcium and Argon. This choice is motivated by the following. Neutron-rich Calcium isotopes have been                 
studied in Ref.~\cite{Hirata} within the relativistic Hartree (RH) theory, a very popular, although not 
microscopic, approach to describe nuclear properties. Neutron and proton radii, charge radii, and 
binding energies are given in that work for several parameter sets of the model. 
At the same time, 
reaction cross section measurements exist for some of those isotopes. This gives us some reference point to    
compare our predictions with as we move continuously from our nuclear density predictions to the reaction cross section calculations. 

As before, 
our nuclear densities are extracted assuming Thomas-Fermi functions as in 
Eq.~(9) together with our microscopic equation of state. A comparison with RH predictions is shown in 
Table II. There, for given Z and A, the first set of predictions is from our work whereas the second is from 
Ref.~\cite{Hirata}, specifically the parameter set referred to as ``HS" \cite{HS}, with the 
exception of $^{64}$Ca, for which results were given only with the set ``TS" \cite{TS}.   
(Notice that the TS predictions, shown for the last isotope, are generally smaller than the HS ones, 
which, on the other hand, are very similar to those from models that include non-linearities
in the $\sigma$ potentials \cite{NL}.) 

First, we note that, overall, the predictions are in reasonable agreement with each other. Considering that 
empirical values for nuclear matter saturation properties and the symmetry energy are used to
determine the parameters in the Hartree theory, it appears that our microscopic EoS has reasonable predictive 
power, even when used in a simple liquid drop model. 

The average isospin asymmetry parameter for the isotopes shown in the Table changes from 0 to 0.375, the last being considerably
larger than in lead. 
Generally, the proton radius changes only little with increasing neutron population. For A=40,
the neutron skin is very small and negative, as discussed previously.                        
 Beyond A=48, in our predictions the skin increases continuously with increasing N, 
from -0.046 fm to 0.553 fm. Na isotopes with A from 20 to 32 were studied in 
Ref.~\cite{Suzuki} (also using parametrizations as in Eq.~(10)) and a monotonic increase in the neutron skin thickness
was observed as well. 

The data shown in Fig.~6 are part of 
measurements taken at the National Superconducting Cyclotron Laboratory \cite{Licot} at         
energies/particle between 50 and 70 MeV (depending on the isotope), where silicon detectors were also utilized as targets. 
The predictions we show next are obtained assuming a target of $^{28}$Si and some of the different Calcium 
and Argon 
isotopes as projectiles. (We are limiting our considerations to isotopes with even A and Z, to maximize the validity of the spherical symmetry assumption.) 
At this point we neglect isospin dependence of the NN cross sections because of the observations made 
at the end of the previous section. 

Figure 6 shows reaction cross section predictions and data for Calcium isotopes (left frame) and Argon (right frame).
The various curves are obtained with: free-space NN cross sections (pink, dotted); medium modified NN cross 
sections as in method I (dash-dotted, blue); medium modified NN cross sections as in method II (dashed, green); medium modified NN cross sections as in method III (dash-double-dotted, orange). A moderate
average density of 0.068 fm$^{-3}$ (between 1/2 and 1/3 of normal nuclear density) is assumed. 

\begin{figure}[!t] 
\centering 
\vspace*{-1.0cm}
\hspace*{-1.0cm}
\scalebox{0.30}{\includegraphics{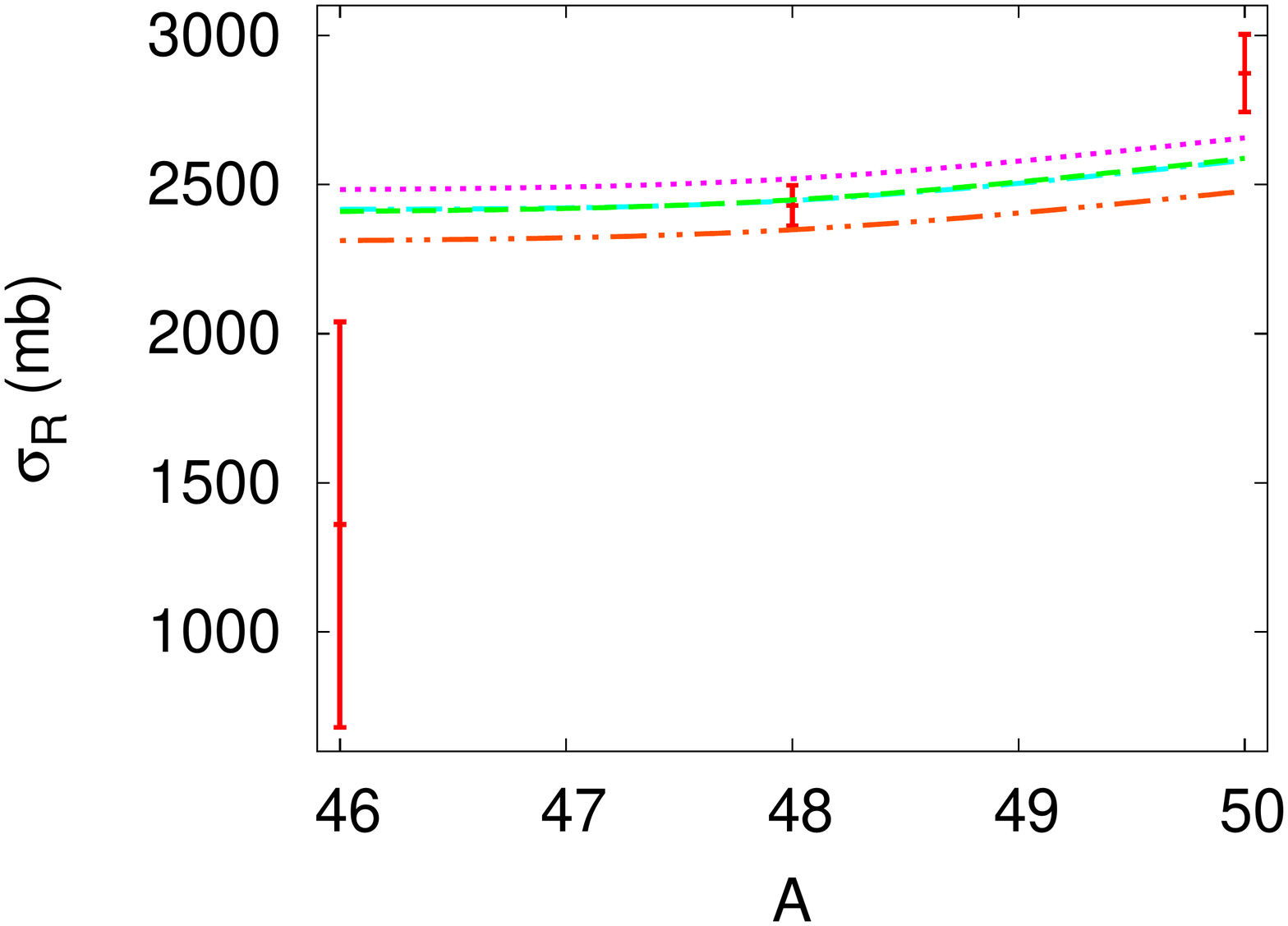}} 
\scalebox{0.30}{\includegraphics{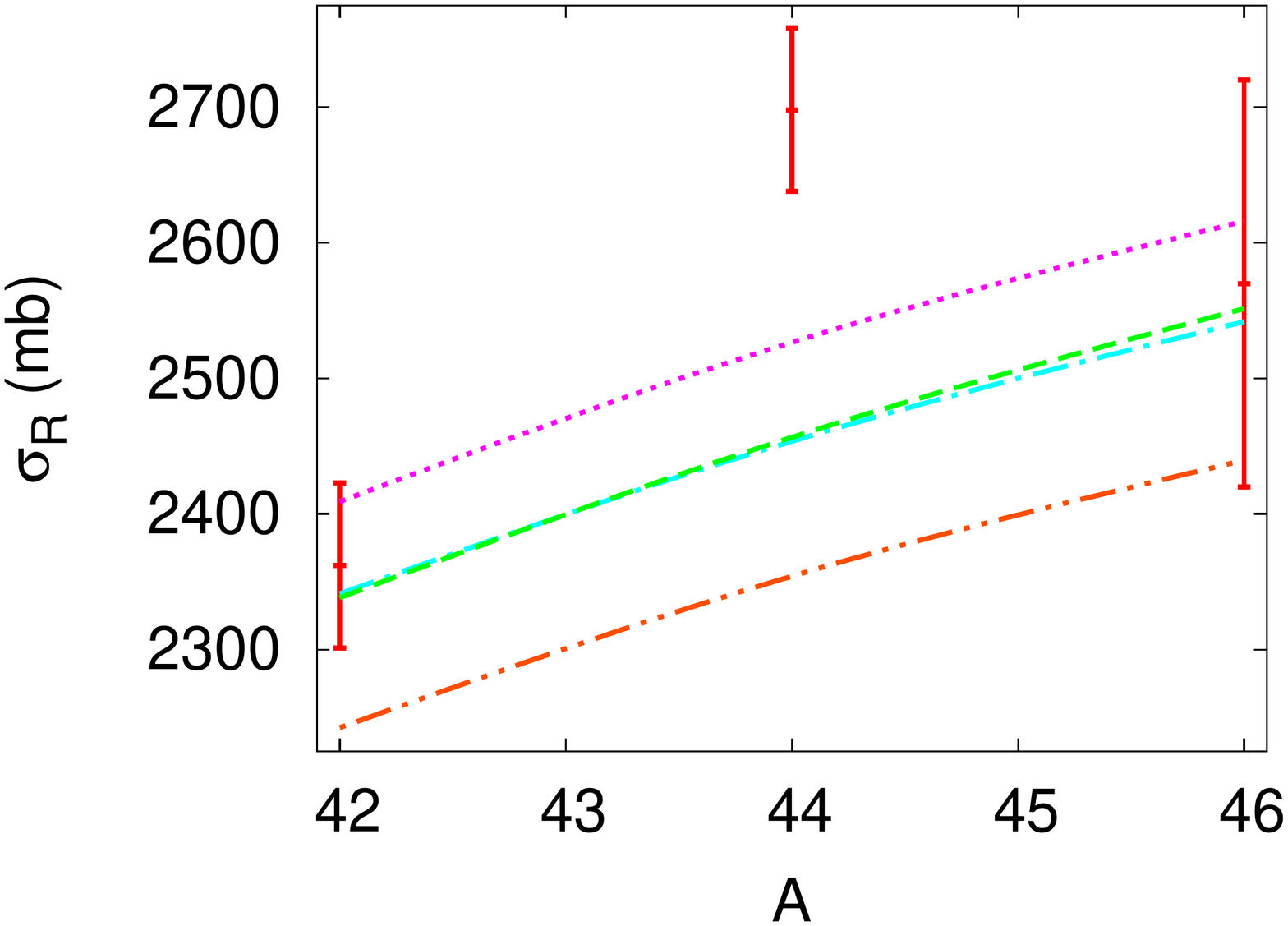}} 
\vspace*{-0.5cm}
\caption{(color online)                                        
Total reaction cross sections as a function of the mass number of Calcium isotopes (left frame) 
and Argon isotopes (right frame). Dotted (pink) line: free-space NN cross sections; Dash-dotted (blue) line:
in-medium cross sections from method I; dashed (green) line: in-medium cross sections from method II;  
dash-double-dotted (orange) line: in-medium cross sections from method III.
Data from Ref.~\cite{Licot}. 
} 
\label{six}
\end{figure}

Although they cannot be discerned at the level of
precision of these data, medium effects can be significant. 
It is important for theorists to agree on accurate calculations of medium effects before
other issues can be                                                    
addressed reliably (such as the nature of ``exotic" nuclear matter distribution).

\begin{table}                
\centering \caption                                                    
{ Some nuclear properties for the neutron-rich isotopes considered in this work. For each set of (Z,A) values, 
 the first row contains predictions from this work, whereas the second     
is from Ref.~\cite{Hirata}. The items shown in the table are the (point) neutron, (point) proton, and charge r.m.s. radii,
and the energy per particle. 
} 
\vspace{5mm}

\begin{tabular}{|c|c|c|c|c|c|}
\hline
$Z $ & $A $ & $r_p(fm)$ & $r_n(fm)$ & $r_{ch}(fm)$ & e/A (MeV) 
\\                                                  
\hline
       20 & 40 &  3.433 & 3.387 &  3.525 & -7.68 \\ 
          &    &  3.387 & 3.332 &  3.483 & -6.51 
\\ 
       20 & 46 &  3.448 & 3.555 &  3.539 & -7.91 \\ 
          &    &  3.374 & 3.535 &  3.470 & -6.76 
\\ 
       20 & 48 &  3.446 & 3.545 &  3.538 & -7.82 \\ 
          &    &  3.375 & 3.588 &  3.471 & -6.86 
\\ 
       20 & 50 &  3.479 & 3.702 &  3.571 & -7.78 \\ 
          &    &  3.394 & 3.733 &  3.490 & -6.72 
\\ 
       20 & 52 &  3.506 & 3.790 &  3.596 & -7.67 \\ 
          &    &  3.414 & 3.854 &  3.508 & -6.61 
\\
       20 & 54 &  3.535 & 3.845 &  3.625 & -7.53 \\ 
          &    &  3.440 & 3.929 &  3.534 & -6.42 
\\
       20 & 60 &  3.616 & 4.080 &  3.704 & -7.04 \\ 
          &    &  3.514 & 4.144 &  3.606 & -6.05 
\\
       20 & 64 &  3.663 & 4.216 &  3.750 & -6.67 \\ 
          &    &  3.348 & 3.878 &  3.445 & -7.12 
\\ 
\hline

\end{tabular}
\end{table}

\section{Conclusions}                                                                  

In closing, 
we stress that one of the main features of our work is 
a seamless and parameter-free pipeline (from the effective NN interaction to the EoS, to nuclear densities, to in-medium NN cross sections, to reaction cross sections) originating 
from one single realistic NN potential \cite{bonn}.                            
For the purpose of this sensitivity study, 
we adopted a reaction model that has been traditionally used due to its convenience and simplicity. 
In the near future, we plan to remove some of the approximations inherent to the optical limit of the Glauber model \cite{Abu}.

We have performed sensitivity tests with the total reaction cross section involving 
stable (although mildly isospin asymmetric) nuclei, such as $^{208}$Pb. This observable has the
potential to discriminate among different EoS yielding different predictions for the neutron distribution
of neutron-rich nuclei. 
 
We explored the impact of applying 
medium effects  on the NN cross sections. Taking as a reference point 
standard data sets with stable nuclei such as those by Kox {\it et al.} \cite{Kox}, we see that 
error bars can be as small as 6-7\% and occasionally as large as 20\% or more. Overall, this level of precision
would make it difficult 
to resolve the model dependence of the reaction cross section with respect to medium effects. Nevertheless,                 
theorists should feel encouraged to find better agreement concerning the best definition and calculation 
of effective in-medium NN cross sections.

In a problem as complex as the one of heavy-ion reactions, where several ingredients are necessary
and none of them is free of uncertainty, 
one has to be very selective of the appropriate ``laboratory". 
 Reactions with exotic nuclei at high energy (as those measured, for instance, in Ref.~\cite{Suzuki} at 950A MeV), where medium effects can be expected to be negligible, are best suited to probe matter distribution beyond stability. Once reliable information on the structure is available,                             
that information can be used in calculations of 
lower-energy reactions where the focus can be placed on the density-dependence of the nuclear interaction. 

As noted in the Introduction, 
the isospin degree of freedom plays an important role in heavy-ion collisions, especially  the 
transport of neutrons and protons.  
It was found \cite{Li05} that in-medium NN cross sections have significant influence
on the isospin transport and the nucleon transverse flow at intermediate energy. 
Although here we are not dealing with a transport model, 
the transparency, written as in Eq.~(8), allows for a natural separation of the $nn$, $pp$, and $np$ 
cross sections. Sensitivity to such separation, if any, would result from different             
probabilities (of the different types of collisions) 
in the overlap region of isospin asymmetric colliding ions. We explored whether such differences
can be detected by the reaction cross section and concluded that they cannot.

Ultimately, the possibility of exploiting this and other reaction observables to constrain nuclear 
properties and/or medium modifications of the nuclear force 
rests on the availability of 
large systematics and good-precision data. These are precisely among the major
goals of upcoming experimental efforts in nuclear physics.

\section*{Acknowledgments}
Support from the U.S. Department of Energy under Grant No. DE-FG02-03ER41270 is 
acknowledged.                                                                           

\end{document}